\documentclass[aps,prx,twocolumn,showpacs,superscriptaddress,longbibliography]{revtex4-2}
\usepackage{comment}
\usepackage{amsmath,amssymb,latexsym,mathrsfs}
\usepackage{graphicx} 
\usepackage{color}
\usepackage{accents}
\usepackage{float}
\usepackage{gensymb}
\usepackage{tikz-feynman}
\usepackage[normalem]{ulem}
\usepackage[breaklinks=true,colorlinks,citecolor=blue,linkcolor=blue,urlcolor=blue]{hyperref}
\begin{document}
\newcommand{\be}{\begin{eqnarray}}
\newcommand{\ee}{\end{eqnarray}}
\newcommand{\lb}{\label}
\newcommand{\bs}{\boldsymbol}
\newcommand{\cb}{\color{blue}}
\newcommand{\st}{\sout}
\newcommand{\ssp}{{\sigma \sigma^\prime}}
\title{Non-orthogonal spin-momentum locking}  
\author{Tugrul Hakioglu $^{\dag}$}
\email{hakioglu@itu.edu.tr}
\affiliation{Energy Institute and Department of Physics, Istanbul Technical University \\ 
Maslak 34469, Istanbul, Turkey}
\author{Wei-Chi Chiu}
\thanks{These authors contributed equally.}
\affiliation{Department of Physics, Northeastern University, Boston, MA 02115, USA} 
\author{Robert S. Markiewicz}
\affiliation{Department of Physics, Northeastern University, Boston, MA 02115, USA}
\author{Bahadur Singh}
\affiliation{Department of Condensed Matter Physics and Materials Science, Tata Institute of Fundamental Research, Colaba, Mumbai 400005, India}
\author{Arun Bansil}
\affiliation{Department of Physics, Northeastern University, Boston, MA 02115, USA} 
\begin{abstract} 
Spin-momentum locking is a unique intrinsic feature of strongly spin-orbit coupled materials and a key to their promise of applications in spintronics and quantum computation. Much of the existing work, in topological and non-topological pure materials, has been focused on the orthogonal locking in the vicinity of the $\Gamma$ point where the directions of spin and momentum vectors are locked perpendicularly. With the orthogonal case, enforced by the symmetry in pure systems, mechanisms responsible for non-orthogonal spin-momentum locking (NOSML) have drawn little attention, although it has been reported on the topological surface of $\alpha$-$Sn$. Here, we demonstrate that, the presence of the spin-orbit scattering from dilute spinless impurities can produce the NOSML state in the presence of a strong intrinsic spin-orbit coupling in the pristine material. We also observe an interesting coupling threshold for the NOSML state to occur. 

The relevant parameter in our analysis is the deflection angle from orthogonality which can be extracted experimentally from the spin-and-angle-resolved photoemission (S-ARPES) spectra. Our formalism is applicable to all strongly spin-orbit coupled systems with impurities and not limited to topological ones. The understanding of NOSML bears on spin-orbit dependent phenomena, including issues of spin-to-charge conversion and the interpretation of quasiparticle interference (QPI) patterns as well as scanning-tunneling spectra (STS) in general spin-orbit coupled materials.
\end{abstract}
\maketitle 
\section{Introduction}
Spin-momentum locking (SML) occurs commonly in spin-orbit coupled low dimensional materials with or without topological bands\cite{winkler2003spin,dyakonov2017spin,doi:10.1146/annurev-conmatphys-062910-140432,RevModPhys.88.021004}. Its telltale signatures involve forbidden backscattering \cite{franz2013topological,ortmann2015topological} from non-magnetic impurities (no ‘U-turn’) and enhancement of weak antilocalization effects
\cite{WAL}. SML enables electrical control of spin polarization in nonequilibrium transport and thus plays a key role in spintronics and spin-based quantum information sciences applications \cite{awschalom2013semiconductor,dyakonov2017spin} in the capability to drive a spin-polarized current with polarization perpendicular to the current density \cite{PhysRevLett.105.266806,PhysRevLett.105.066802,PhysRevB.82.155457}.

The orthogonal SML (OSML)--see Fig.\ref{osml_vs_nosml_main_figure}.a, is common in materials exhibiting SML\cite{winkler2003spin,dyakonov2017spin,doi:10.1146/annurev-conmatphys-062910-140432}. The OSML state is the result of in-plane Rashba spin-orbit coupling (SOC) observed first time on the Au (111) surface long before the topological materials were discovered\cite{PhysRevLett.77.3419}.  
In topological insulators, OSML with a $\pi$-Berry phase is an essential feature of the surface electron bands\cite{RevModPhys.82.3045}. It has been utilized in the electrical detection of magnon decay\cite{Jiang:2016aa}. Despite its broad presence in strongly spin-orbit coupled materials, violations of the SML has been seen in real materials. In such cases, spin and momentum are weakly unlocked within a narrow range of angular deviations constrained by the crystal symmetries. An S-ARPES study of the Au/Ge (111) surface revealed such examples\cite{PhysRevLett.108.186801} and similar effects have been reported in high-temperature superconductors\cite{Gotlieb1271}. In certain topological insulators the spin wiggles around the Fermi surface due to the hexagonally warped Fermi surface but respects the OSML\cite{PhysRevLett.103.266801}. Deviations from the orthogonal picture were also observed experimentally in the $Bi_{2-y}Sb_yTe_xSe_{3-x}$ family\cite{PhysRevLett.107.207602} as shown in the Fig.\ref{osml_vs_nosml_main_figure}.b. We can call these weak violations from the orthogonally locked state as type-I violations. It has been shown that high-order corrections to the theoretical ${\bs k}.{\bs p}$ Hamiltonian can induce deviations from the orthogonal picture\cite{PhysRevB.84.121401}. Many body interactions also cause similar effects, as the electron-phonon interaction in this material\cite{PhysRevB.89.075425,PhysRevLett.108.185501,PhysRevLett.108.187001,Heid:2017aa} was recently studied in this context\cite{PhysRevB.97.245145,PhysRevB.100.165407}. The triple and septuple windings of the spin vector have also been studied theoretically as violations of the OSML\cite{PhysRevLett.124.237202,PhysRevB.102.115437}.

Another type of deviation from the perfect OSML state is not in the locking of the spin and momentum but in their orthogonality, i.e. the non-orthogonal spin-momentum locking (NOSML) as illustrated in Figs.\ref{osml_vs_nosml_main_figure}.c and d (called as the type-II violations of the OSML). Such a state has been reported on the topological surface of strained $\alpha$-$Sn$ \cite{PhysRevB.97.075101,PhysRevLett.111.157205}.  
Here, S-ARPES and Mott polarimetry reveal the presence of   
a radial component of the spin (Fig\ref{osml_vs_nosml_main_figure}.c) with a significant inward deviation of $\Phi_0-90^\circ \simeq 20^\circ$ on a circular Fermi surface\cite{deviation_angle_alpha_Sn}. The out-of-plane spin $S_z$ is observed to vanish in conformity with the absence of the out-of-plane SOC. Note that $\alpha$-$Sn$ is inversion symmetric in unstrained and strained phases~\cite{PhysRevLett.111.157205,PhysRevB.95.161117,PhysRevB.95.201101,PhysRevLett.111.216401,PhysRevLett.116.096602,PhysRevLett.118.146402,PhysRevB.98.195445,PhysRevB.76.045302}. The authors of Ref.\cite{PhysRevB.97.075101} also point at the presence of electron-impurity interaction through an analysis of the electronic self-energy.

This last observation is of key importance in our theory of the NOSML. Our approach is not limited to topological surface states but addresses NOSML as a general phenomenon in materials with strong SOC. The presence of inversion and time-reversal symmetries substantially constrains the Hamiltonian for treating non-interacting bands. Origin of the NOSML lies beyond the realm of warped electronic bands and details of the lattice structure are not important for generating this effect. Indeed, the OSML state is strictly enforced in pure materials due to the $C_{\infty v}$ symmetry. We therefore study impurity effects in this article as the source of NOSML.  

\section{The Theory of Interacting Spin}
Our starting point is the time-reversal invariant Hamiltonian in the pseudo-spin $\vert \bs k \sigma\rangle$ basis ($\hbar=1$)\cite{winkler2003spin,bir1974symmetry,voon2009k}:
\be
{\cal H}_0= (\xi_{k}-\mu) \sigma_0 + \mathfrak{g}_{\bs k}.{\boldsymbol \sigma}
\lb{linear_spect}
\ee

where $\bs \sigma=(\sigma_x,\sigma_y,\sigma_z)$ is the pseudospin representing the spin-orbit coupled total angular momentum states\cite{winkler2003spin}, ${\bs k}=(k_x,k_y)=k(\cos\phi,\sin\phi)$ is the electron wavevector relative to the Dirac point at $\bs k=0$ and $\xi_{k}$ and $\mu$ are the spin-independent and isotropic bare electron band and the chemical potential, respectively. The Hamiltonian in Eq.(\ref{linear_spect}) is the most basic Hamiltonian in spintronics as well as topological surfaces. A pair of such Hamiltonians can be used to model states in Dirac and Weyl semimetals as well as Rashba type interface states. The spin-orbit vector $\mathfrak{g}_{\bs k}$ is normally composed of an in-plane component ${\bs g}_{\bs k}=g_0 \,\hat{\bs z}\times {\bs k}$ with $g_0$ as the Rashba type in-plane SOC and $\hat{\bs z}$ as the surface unit normal vector to the $xy$ plane defined by the $\bs k$ vector, and an anisotropic out-of-plane component ${\bs g}_{\perp \bs k}$. We also represent the in-plane and out-of-plane components of the spin as well as the self-energy vectors below using the  same notation. The ${\bs g}_{\perp \bs k}$ as well as the out-of-plane component of the spin are zero in our case due to the azymuthal rotational symmetry. The eigenstates $\vert \bs k \lambda\rangle$ of Eq.(\ref{linear_spect}), where $\lambda=\pm$ is the spin-orbit band index, include the chiral spin-1/2 state not only attached to the dominant $\vert p_z\rangle$ orbitals as considered conventionally, but also the in-plane orbitals $\vert p_x \rangle$ and $\vert p_y \rangle$. The role played by the in-plane orbitals is strongly material dependent, which has been demonstrated experimentally\cite{Cao:2013aa} and theoretically\cite{PhysRevLett.111.066801}. It is known that these effects do not violate the OSML\cite{RevModPhys.82.3045,PhysRevLett.103.266801}. For our purposes in this work we ignore these in-plane orbitals and consider that the orbital texture is solely determined by the out-of-plane $\vert p_z\rangle$ orbitals. Further discussion on this point is made in Section.V. With this considered, the pseudospin is given by $\langle \bs J \rangle=\langle \bs k \lambda\vert {\bs \sigma} \vert \bs k  \lambda\rangle=(\lambda/2)\hat{\mathfrak{g}}_{\bs k}$, where $\hat{\mathfrak{g}}_{\bs k}$ is the unit spin-orbit vector, which coincides with the actual spin $\bs S_\lambda(\bs k)=(\lambda/2)\hat{\mathfrak{g}}_{\bs k}$. Since the in-plane component ${\bs g}_{\bs k}$ of the spin-orbit vector $\mathfrak{g}_{\bs k}$ is perpendicular to $\bs k$, the spin $S_\lambda(\bs k)$ is locked orthogonally to the electron momentum $ \bs k$ throughout, yielding the OSML.   

The Eq.(\ref{linear_spect}) is clearly insufficient to describe all strongly spin-orbit coupled surfaces and additional terms may be present due to the symmetries. For instance, the cubic Dresselhaus SOC is present in the absence of inversion symmetry in ordinary semiconductors\cite{winkler2003spin}. Its realization in $Bi_2X_3$ type strong topological insulators results in the hexagonal warped Fermi surfaces but the OSML is still respected\cite{PhysRevLett.103.266801}.       
\begin{figure}
\includegraphics[scale=0.4]{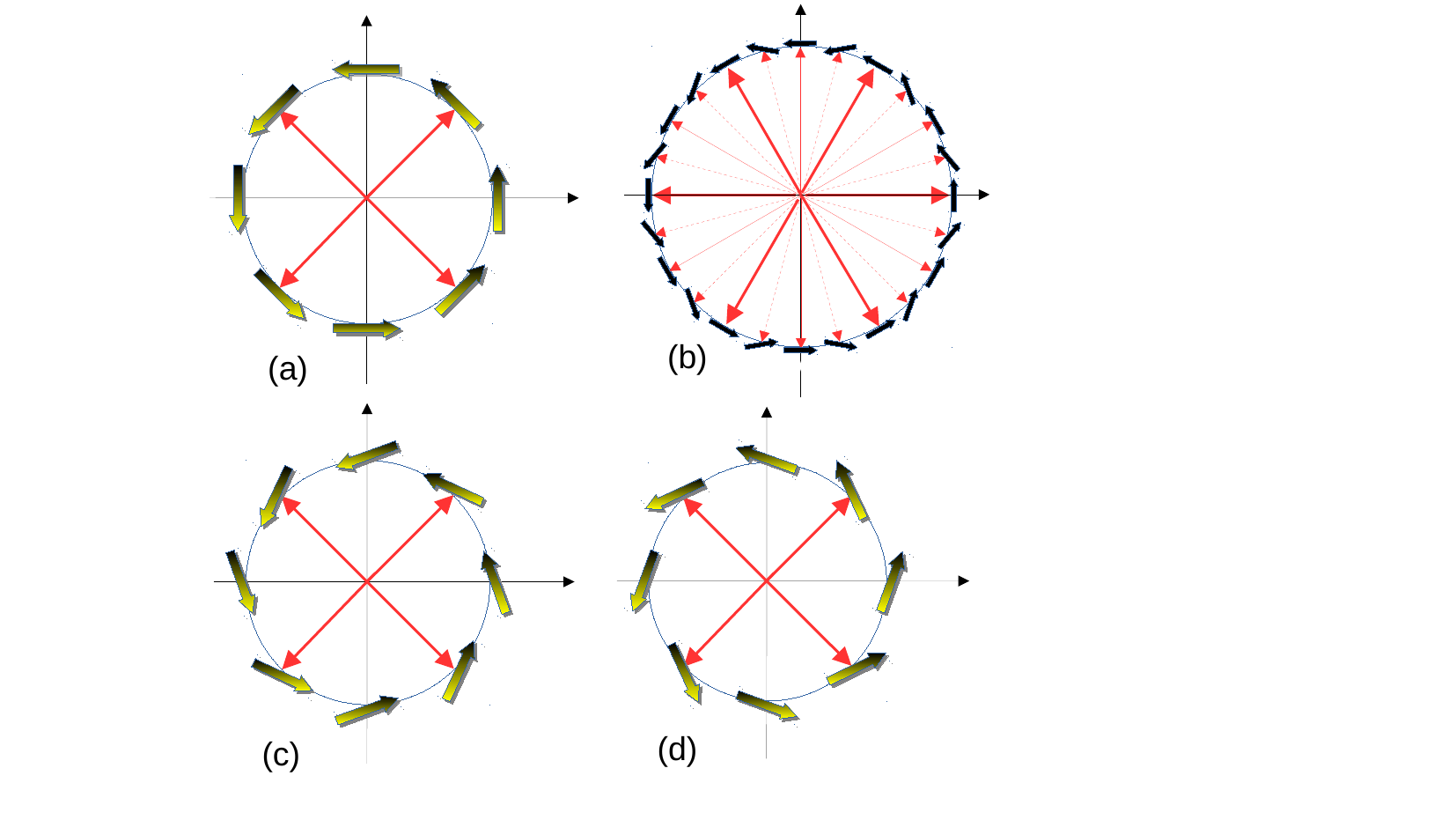}    
\caption{A schematic of various planar spin-momentum locking cases for a chiral band. (a) The OSML. (b) Weakly unlocked case of $Bi_2Se_3$ and $Bi_2Te_3$. The weak out-of-plane component is not shown. (c, d) NOSML with $\delta < 0$ (c) and $\delta  > 0$ (d). $\delta$ is defined in Eq.(\ref{nosml_angle}).}
\label{osml_vs_nosml_main_figure}
\end{figure}
In few other cases, the anisotropy may persist down to the $\Gamma$ point\cite{PhysRevB.84.195425}. Note that, NOSML is ideally an isotropic effect and anisotropic warping in the band structure may hinder its observation. Therefore, we ignore the anisotropy and examine rotational symmetry allowed Hamiltonians in the study of the NOSML state. The minimal Hamiltonian which can yield a NOSML state is\cite{winkler2003spin,bir1974symmetry,voon2009k} 
\be 
{\cal H}_1=\gamma \, {\bs k}.{\bs \sigma}~.
\lb{IS_broken}
\ee 
This Hamiltonian is known to be present in the Kane model between the $\Gamma_{7c}$ and $\Gamma_{6c}$ bands of zinc-blende structures\cite{PhysRevB.44.9048,winkler2003spin,PhysRevB.91.165435}. The total Hamiltonian ${\cal H}_0+{\cal H}_1$ is equivalent to ${\cal H}_0$ by a complex rotation of the spin-orbit constant, and effective spin-orbit coupling is defined as ${\bs g}_{\bs k}=g_0 \,\hat{\bs z}\times {\bs k}+\gamma \bs k$. The energy spectrum is linear and spin-momentum pair is locked non-orthogonally at $\Phi_0=\pm (\pi/2 + tan^{-1} \gamma/g_0)$ with the $\pm$ describing the upper and the lower Dirac cones. While, Eq.(\ref{IS_broken}) can easily accommodate a non-orthogonal state of the spin and momentum in zinc-blende structures\cite{PhysRevB.91.165435}, it is not applicable when the inversion symmetry holds since that requires $\gamma=0$. This prompts us to think that the NOSML in inversion symmetric systems may have its origin fundamentally beyond the class of symmetry allowed single particle Hamiltonians.

Here we demonstrate that, the impurities in the real materials may provide a striking clue for the general source of the NOSML. It is known that the electron-impurity interaction, combined with the strong SOC, gives rise to the spin-orbit scattering in addition to the scalar scattering channels, which then leads to a number of observable transport phenomena. These are linearly dependent on the spin-orbit scattering strength\cite{ma10070807} such as corrections in the momentum and spin relaxation, spin-dependent diffusion, weak localization/antilocalization\cite{dyakonov2017spin} and anomalous spin-texture\cite{Bhattacharyya2019}. The spin-orbit scattering between the impurity and the electron bands also provides a platform for NOSML and this is the main focus of this work. 

In this work, we use the interacting Green's function formalism for the renormalized spin\cite{PhysRevB.100.165407}. In this approach the spin is given by      
\be
{\bs S}_\lambda({\bs k})=\frac{\lambda}{2}\hat{\bs G}({\bs K}^*)
\lb{spin_1.3}
\ee
Here, the $*$ indicates that ${\bs S}_\lambda({\bs k})$ is calculated at the physical energy pole position $E^*=E_{\lambda k}$ of the full Green's function\cite{PhysRevB.100.165407}. ${\bs G}({\bs K}^*)={\mathfrak g}_{\bs k}+{\bs \Sigma}({\bs K}^*)$, with ${\bs K}^*=({\bs k},i E^*)$ in the Matsubara Green's function formalism, is the renormalized spin-orbit vector where $\hat{\bs G}={\bs G}/\vert {\bs G}\vert$ is its unit vector. Here it is crucial to note that, ${\bs G}$ enters as a simple sum of the spin-orbit vector ${\mathfrak g}_{\bs k}$ of the non-interacting structure and the interactions represented by the spin-dependent self-energy ${\bs \Sigma}$ (SDSE). The ${\mathfrak g}_{\bs k}$ is aligned perpendicularly to the momentum. The renormalized spin-orbit vector ${\bs G}$ however, develops a non-orthogonal component as a result of the interactions. We recently demonstrated this theoretically using the electron-phonon interaction and the Fermi surface warping yielding six-fold symmetric type-I violations of the OSML  in the topological insulator $Bi_2Se_3$\cite{PhysRevB.97.245145,PhysRevB.100.165407} [see Fig\ref{osml_vs_nosml_main_figure}.(b)]. Here, we demonstrate that, the electron-impurity interaction can have a similar consequence without the need of a warped Fermi-surface yielding a non-orthogonally locked configuration of the spin and momentum [Fig\ref{osml_vs_nosml_main_figure}.(c) and (d)].   

The spin-dependent-self-energy (SDSE) vector ${\bs \Sigma}({\bs K}^*)$ represents the  impurity average of the microscopic scattering events between the electron and the impurity (see Appendix A). The full spin-neutral and the spin-dependent parts of the self energies can be combined in the pseudospin matrix form as,      
\be
\underline{\Sigma}({\bs K})=\Sigma_0({\bs K}) \sigma_0 +{\bs \Sigma}({\bs K}).{\bs \sigma}
\label{SE_1}
\ee 
where $\Sigma_0$ is the spin neutral self-energy (SNSE) and ${\bs \Sigma}=(\Sigma_x,\Sigma_y,\Sigma_z)$ is the SDSE as introduced before. The total change in the spin between this interacting model and the non-interacting one is expectedly decided by the SDSE which is given by 
$\Delta {\bs S}_\lambda({\bs k})=(\lambda/2)\,[\hat{\bs G}^*({\bs k})-\hat{\mathfrak g}_{\bs k}]$. Here we took the difference of two cases with and without interactions using Eq.(\ref{spin_1.3}). In the weak interaction limit\cite{PhysRevB.100.165407} this takes an elegant form with the leading term      
\be
\Delta {\bs S}_{\lambda}({\bs k})\simeq \frac{\lambda}{2} \, \frac{\Sigma^*_k}{\vert {\mathfrak g}_{\bs k}\vert}  \, \hat{\bs k} 
\label{spin_1.6}
\ee 
where $\Sigma^*_k={\bs \Sigma}({\bs K}^*). {\hat{\bs k}}$ is the component of the SDSE along the momentum. The Eq.(\ref{spin_1.6}) states that the interactions can cause both type-I and type-II violations of the OSML. Since there are strong symmetry considerations in the pure crystal structure, the generation of a finite $\Sigma^*_k$ is not trivial. In this work, we study the electron-non-magnetic impurity scattering as a new mechanism for the non-orthogonally locked type-II state  as shown in Fig.\ref{osml_vs_nosml_main_figure} (c) and (d).    

\section{The spin-orbit impurity scattering}
Whatever the mechanism is, the formalism in Eqs. (\ref{spin_1.3}-\ref{spin_1.6}) hinges upon an accurate model for the self-energy in Eq.(\ref{SE_1}). The electron-impurity scattering is represented as a scattering potential $V_{e\,i}^{(j)}=V_{0}^{(j)}+V^{(j)}_{so}$ the spin independent and spin-orbit scattering parts are given by  $V_{0}^{(j)}$ and $V^{(j)}_{so}$ respectively. In the notation, the superscript refers to the $j$'th impurity (see Appendix B). The spin-orbit coupling in the pristine sample is assumed to be sufficiently strong compared to the electron-impurity interaction. By this approach a simple picture can be obtained where the leading order contribution to NOSML can be isolated from the other secondary effects. The scattering matrix between the initial $\vert i\rangle=\vert {\bs k} \, \sigma\rangle$ and the final $\vert f\rangle=\vert {\bs k}^\prime \, \sigma^\prime\rangle$ states is $T_\ssp (\bs k,\bs k^\prime)=\sum_j \,\langle {\bs k}^\prime \, \sigma^\prime \vert V_{e\,i}^{(j)} \vert {\bs k} \, \sigma\rangle$ given in the Born approximation by\cite{mahan2012many,Zhong:2017aa,fischetti2016advanced,PhysRevLett.119.117001,dyakonov2017spin,ma10070807,mott1949theory,landau2013quantum,burke2012potential,PhysRevLett.95.166605,PhysRevB.95.115307,
PhysRevB.86.125303,PhysRevB.80.245439} and Appendix B by, 
\be 
T_\ssp (\bs k,\bs k^\prime)= \sum_j\, e^{i(\bs k-\bs k^\prime).\bs R_j} ~ t_\ssp^{(j)}(\bs k,\bs k^\prime)
\lb{T_kkp}
\ee
where the exponential phase factor accounts for the impurity scattering phase shifts occuring at random centers $\bs R_j$ and $t_\ssp^{(j)}(\bs k,\bs k^\prime)$ is the scattering amplitude of the electron off the $j$'th impurity from the initial to the final state which can be derived microscopically once the impurity-electron scattering potential is known. We will assume that there is only one kind of impurity and drop the $j$ index in $t_{\sigma \sigma^\prime}^{(j)}$. The scattering of an external spinless impurity with the electron under the influence of the spin-orbit coupling is an old textbook problem which has been studied before\cite{landau2013quantum}. The effective interaction is basically a superposition of two independent parts. The first part is a spin independent channel contributing to the momentum distribution and relaxation. The second part has been shown to arise as a result of the interaction between spin-orbit coupled electrons and the spinless impurity. The scattering matrix is then given by\cite{PhysRevB.95.115307} (see Appendix B),   
\be 
{\underline t}(\bs k,\bs k^\prime)=\, a_0 \,\sigma_0+ c_0\,\hat{\bs k}\times \hat{\bs k}^\prime . \, {\bs \sigma}
\lb{t_0.1}
\ee
where the $t_\ssp$ in Eq.(\ref{T_kkp}) corresponds to the matrix element of the Eq.(\ref{t_0.1}) with the spin indices $\sigma, \sigma^\prime$. The first term describes the spinless scattering with $a_0$ as the scattering strength and the second term is the spin-orbit scattering with the strength $c_0$. These coefficients are generally functions of $\bs k$, $\bs  k^\prime$ as well as the details of the microscopic electron-impurity interaction   
\cite{dyakonov2017spin,mott1949theory,landau2013quantum,burke2012potential,PhysRevLett.95.166605} (see Appendix B). 

We further assume dilute impurity limit $n_i \ll \lambda_F^{-3}$ where $n_i$ is the impurity concentration and $\lambda_F$ is the Fermi wavelength of the scattered electrons. In this limit, we neglect the interference between multiple scattering events.

\section{The Self-Energy due to the Spin-Orbit Scattering}
\subsection{The spin-independent self-energy}
The OSML is strictly enforced by the $C_{\infty v}$ symmetry near the $\Gamma$ point in pure crystals\cite{PhysRevB.76.073310,PhysRevB.68.165416}. In strongly spin-orbit coupled materials, deviations from this orthogonal picture requires a sufficient impurity coupling and the renormalization of the Bloch states. Including the impurity scattering perturbatively, the effect vanishes in the first order of the perturbation since at this level the electron self energy averages out to zero over the impurities (see Appendix A). Here, the NOSML emerges beyond the second order in the electron self energy and this includes the renormalization of the Bloch states. The Feynman diagrams of the Green's functions and the self energies are summarized in Fig.(\ref{Feynman_1}) of the Appendix A. 

Another point to stress is that, the NOSML can be concealed by warping or other anisotropy effects. To keep the formulation at a fundamental level, we limit ourselves to the case when such phenomena are absent or sufficiently weak and consider an isotropic band $\xi_{k}= k^2/(2m)$ near the $\Gamma$ point. The full impurity averaged self-energy in Eq.(\ref{SE_1}) is defined as \cite{mahan2012many,Zhong:2017aa,RevModPhys.78.373}, 
\be
&&\underline{\Sigma}(\bs K)= \frac{n_i}{2} \, \int \frac{d{\bs k^\prime}}{(2\pi)^2} \nonumber \\
&&\times \sum_\lambda \underline t({\bs k},{\bs k}^\prime)\,[1+\lambda \, \hat{\bs G}(\bs k^\prime,E).\underline{\bs \sigma}]\,{\cal G}_{\lambda}(k^\prime,E)\,
\underline t({\bs k}^\prime,{\bs k}) ~~~~~~~~
\lb{Sigma_matrix_1}
\ee
The ${\cal G}_\lambda(k,E)=1/(E-E_{\lambda k})$ is the Green's function of the eigenband with index $\lambda$ and  $E_{\lambda k}=\tilde{\xi}_{k}+\lambda \vert {\bs G}({\bs k},E)\vert$ as the renormalized energy band with $\tilde{\xi}_{k}=\xi_{k}+\operatorname{Re}\{\Sigma_{0}\}$ and ${\bs G}=\mathfrak{g}_{\bs k}+{\bs \Sigma}$. The $n_i$ dependence in Eq.(\ref{Sigma_matrix_1}) comes from the averaging over the random impurity positions $\bs R_i$ as given in Eq.(\ref{T_kkp}) and shown in the Appendix A. We note that, the dependence of the self energy on the impurity concentration in Eq.(\ref{Sigma_matrix_1}) is not linear due to the non-linear dependence in the renormalized spin-orbit vector ${\bs G}$ on the self-energy. Furthermore, these equations can be obtained from our more general theory of the surface electrons interacting with the lattice excitations studied in Ref.\cite{PhysRevB.100.165407} when the phonon excitation energy vanishes in the static limit.

The spin-independent and spin-dependent parts of Eq.(\ref{Sigma_matrix_1}) are given by,  
\be 
\Sigma_{0} =\operatorname{Tr}\{\underline{\Sigma}\}/2 \,,\qquad {\bs \Sigma}=\operatorname{Tr}\{\underline{\Sigma}~\underline {\bs \sigma}\}/2
\lb{trace_ids}
\ee
We assume that the spin-orbit scattering is weak compared to the spin-independent one. We also neglect the overall phase of the 
${\underline t}(\bs k,\bs k^\prime)$ and assume that $a_0$ is real. The latter can then be directly related to the spin-independent self-energy $\Sigma_0$. Using the Eq's.(\ref{trace_ids}) we have,   
\be
\operatorname{Im}\{\Sigma_0(E)\} \simeq \frac{m}{4} n_i a_0^2 \Biggl(1 -\frac{g_0}{\sqrt{g_0^2+\frac{2}{m} E}}\Biggr)~.                          
\lb{imaginary_Sigma_0.2} 
\ee  
which is related to the life-time $\tau=1/\operatorname{Im}\{\Sigma_0\}$ of the electron momentum  due to its scattering with the impurities. Another importance of this equation is the connection with the experiment, i.e. $\operatorname{Im}\{\Sigma_{0}(E)\}$ can be directly extracted from the experimental quasiparticle momentum distribution\cite{PhysRevB.97.075101}. We will use the $\operatorname{Im}\{\Sigma_{0}\}$ as a phenomenological parameter replacing $n_i$ dependence throughout. 

It will be shown in the next section that the SDSE as found by the second equation in (\ref{trace_ids}) has a different dependence on $n_i$. This is brought by the renormalized spin-orbit vector on the right hand side in Eq.(\ref{Sigma_matrix_1})  which may lead to a critical boundary separating the OSML and the NOSML phases as discussed below.  

\subsection{The spin-dependent self-energy and the NOSML}
We now turn to the spin-dependent component $\bs \Sigma$ in Eq.(\ref{Sigma_matrix_1}), which can be extracted by using the second of the Eqs.(\ref{trace_ids}).  
Since the out-of-plane component of the $\mathfrak{g}_{\bs k}$ is absent due to the rotational symmetry, $\mathfrak{g}_{\bs k}={\bs g}_{\bs k}$ and $\Sigma_z$ is absent. We start by writing ${\bs \Sigma}=(\Sigma_x,\Sigma_y,0)$ in the polar form using the radial ${\hat{\bs k}}$ and the azymuthal ${\hat{\bs g}}_{\bs k}$ unit vectors as  
\be 
{\bs \Sigma}=\Sigma_g \, {\hat{\bs g}}_{\bs k}+\Sigma_k \,{\hat{\bs k}}\lb{polar_SE}
\ee
where $\Sigma_g={\bs \Sigma}.{\hat{\bs g}}_{\bs k}$ and $\Sigma_k={\bs \Sigma}.{\hat{\bs k}}$ are the components of ${\bs \Sigma}$ along the ${\hat{\bs g}}_{\bs k}$ and ${\hat{\bs k}}$ directions respectively and they are scalar functions independent from the direction of ${\hat{\bs k}}$. Using these scalar components is particularly useful in the impurity averaging since $\Sigma_g$ and $\Sigma_k$ are not affected by the scattering directions of the $\bs k$ vector, a crucial factor in the impurity averaging considering the random orientations in each scattering event (see Appendix A). The Eq.(\ref{polar_SE}) is equivalently written as 
\be
\Sigma_x-i\,\Sigma_y=e^{-i(\phi+\pi/2)} ~C_k  \lb{complex_SE}
\ee 
which is a quite convenient way of writing the SDSE since $C_k$ is represented in terms of the scalar components of the SDSE elegantly as $C_k=\Sigma_g +i \,\Sigma_k$. The real part renormalizes the spin-orbit strength since $g_0 k \to g_0 k+\Sigma_g$. This renormalization can be ignored since $g_0 k$ is sufficiently strong. The imaginary part $\Sigma_k$, on the other hand, is an emerging component which is the main cause of the deviation in the spin-momentum locking angle from the orthogonality as shown in Eq.(\ref{spin_1.6}). Using Eq.(\ref{complex_SE}) in Eq.'s (\ref{trace_ids}) and (\ref{Sigma_matrix_1}) we find  
\be
&&C_k= \int \frac{k^\prime d k^\prime}{2\pi}  z_0(k,k^\prime)\,
\, F(k^\prime) 
\lb{Ti_2_R}
\ee
where   
\be 
F(k^\prime)=\sum_\lambda
\frac{g_0 k^\prime+C_{k^\prime}}{\vert {\bs G}(k^\prime,E)\vert} \frac{\lambda}{E-E_{ \lambda k^\prime}}
\lb{g_av}
\ee
which numerically couples $\Sigma_g$ and $\Sigma_k$. Here $z_0(k,k^\prime)$ is a complex scalar function depending on the scattering strengths in the Eq.(\ref{t_0.1}). In the simplest case of constant scattering strengths, $z_0$ is just a complex number.
\begin{figure}[t]
\includegraphics[scale=0.31,angle=0]{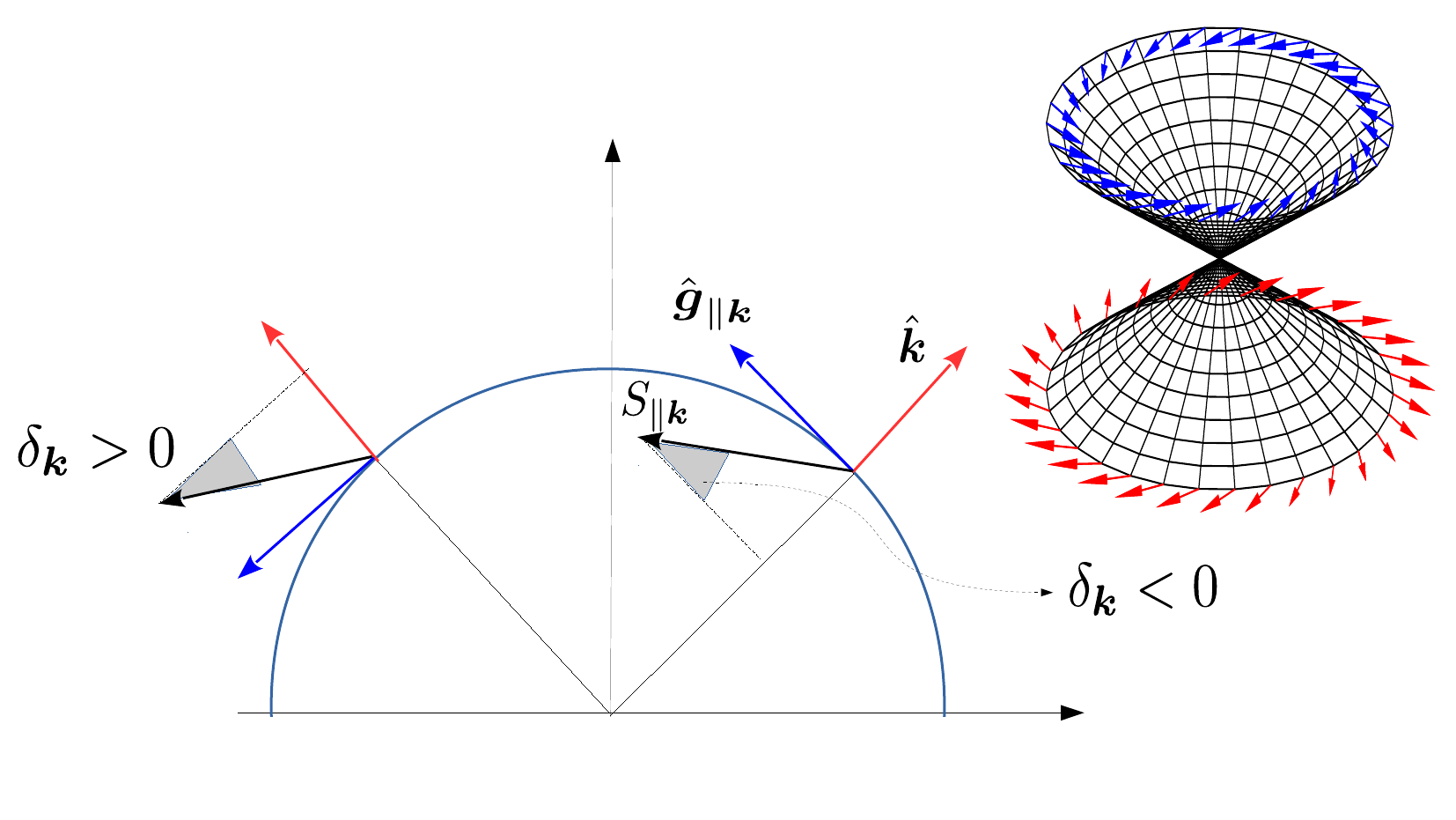} 
\caption{Spin deviation angle $\delta_{k \lambda}$ is illustrated for two different cases $\delta_{k \lambda} <0$ and $\delta_{k \lambda} >0$. The $\delta_{k \lambda}$ has the same sign in the upper and lower Dirac cones which is shown for the $\delta_{k \lambda} >0$ case in the inset.} 
\label{spin_momentum_geometry}
\end{figure}
In order to make a connection with the spin-texture measurements and obtain some quantitative estimates, we now define a microscopic spin-deviation angle $\delta_{\lambda k}$ as illustrated in Fig.(\ref{spin_momentum_geometry}). From the geometry and using Eq.(\ref{spin_1.3}) we find\cite{PhysRevLett.107.207602,PhysRevB.97.245145,PhysRevB.100.165407}  
\be 
\sin \delta_{\lambda k}=\frac{{\bs S}_{\lambda}.\hat{\bs k}}{\vert{\bs S}_{\lambda}\vert} \to \lambda \frac{\Sigma_{k}}{\vert {\bs  G} \vert}  
\lb{nosml_angle}
\ee  
We further identify two cases in Fig.\ref{spin_momentum_geometry} as $\delta_{\lambda k} <0$ and $\delta_{\lambda k} >0$. The Eq.(\ref{nosml_angle}) also yields that the $\delta_{\lambda k}$ in the upper and the lower Dirac cones have opposite signs as  required by the time reversal symmetry and shown by the inlet in Fig.(\ref{spin_momentum_geometry}).   

\begin{figure}[t] 
\includegraphics[scale=0.25,angle=0]{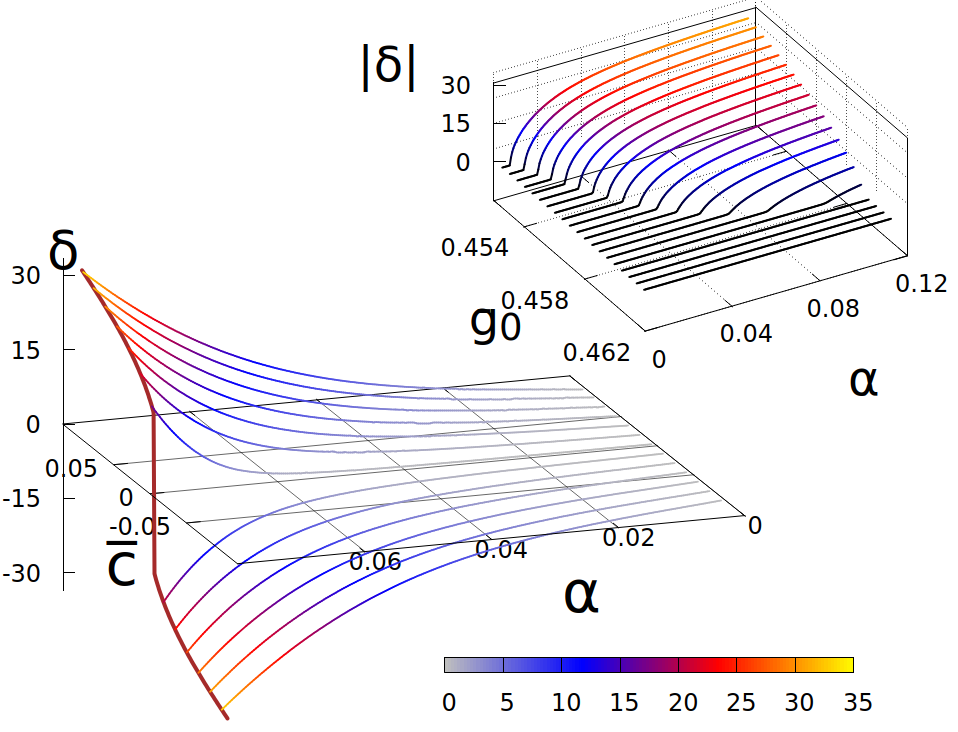}    
\caption{The $\delta_{\lambda k}$ (in degrees) at the Fermi surface for the $\lambda=+$ band in Eq.(\ref{nosml_angle}) using the Eq.(\ref{Ti_2_R}) as the spin-orbit scattering amplitude $\bar{c}$  
and the $\alpha$ are varied at a fixed spin-orbit coupling strength corresponding to $\bar{g}_0=g_0 k_F/E_F \simeq 0.4$. The inset at the top right illustrates a sharp boundary  between the OSML ($\delta=0$) and the NOSML ($\delta \ne 0$) phases determined by the critical values of the $\bar{g}_0$ and $\alpha=\operatorname{Im}{\Sigma}_0/E_F$. The color scale is for the $\vert \delta_{\lambda k} \vert$ and applies to both plots. We used $k_{F}=0.035 \AA^{-1}$ and $E_F=150 meV$ for the normalization\cite{PhysRevB.97.075101}.} 
\label{isotropic_SE_vs_V0_vs_xnu1}
\end{figure}

We now shift our attention to the numerical solution of the Eq.(\ref{Ti_2_R}) which reveals the dependence of the spin-deviation angle $\delta_{\lambda k}$ on the impurity scattering as well as the spin-orbit coupling strengths. We now define a small dimensionless quantity $\alpha=\operatorname{Im}{\Sigma}_0/E_F$ which is linearly dependent on $n_i$. Concerning the solution for the $\delta_{\lambda k}$, we concentrate on the upper Dirac band $\lambda=+$ and solve the Eq.(\ref{Ti_2_R}). It is easy to see that $\delta_{+ k}$ vanishes when $z_0$ is purely real and varies linearly with $\bar c=\operatorname{Im}z_0$ with a steep behavior near $\bar c=0$. The calculated $\delta_{+ k}$ at the Fermi level is shown in Fig.\ref{isotropic_SE_vs_V0_vs_xnu1} as $\bar c$ and $\alpha$ are varied. The inset therein refers to the behaviour when $z_0$ is purely imaginary.         

\section{Discussion and Conclusion}
Due to the conservation of the total angular momentum ${\bs J}={\bs L}+{\bs S}$, the orbital configurations can affect the spin texture\cite{PhysRevLett.111.066801,Cao:2013aa} and, since NOSML is a weak effect due to the small scattering strength $\bar c$, it is important to understand whether the in-plane orbitals $\vert p_x \rangle$ and $\vert p_y \rangle$ can change the observed picture in the Section IV.B. Including the contribution of these in-plane orbitals, the spin-orbital state is given up to the linear order in $k$ by\cite{Cao:2013aa,Dil_2019,PhysRevLett.111.066801},
\be
\vert \vert {\bs k} \lambda \rangle &=&(u_0-\lambda v_1 k) \, (\vert p_z\rangle \otimes \vert \lambda_\phi \rangle) \nonumber \\
&- & \frac{i}{\sqrt{2}}(\lambda v_0 - u_1 k -w_1 k) \, (\vert p_r\rangle \otimes \vert \lambda_\phi \rangle) \lb{spin_orbital_1.1} \\
&+ & \frac{1}{\sqrt{2}}(\lambda v_0 - u_1 k + w_1 k) \, \vert p_t \rangle \otimes \vert \bar\lambda_\phi \rangle \nonumber
\ee
where $u_{0,1},v_{0,1},w_1$ are material dependent coefficients, $\vert \lambda_\phi \rangle=(1/\sqrt{2})[\vert\uparrow\rangle -\lambda i e^{i\phi}\vert \downarrow\rangle]$ is the chiral spin-$1/2$ vortex state, $\vert \bar\lambda_\phi \rangle=\vert (-\lambda)_\phi \rangle$ and $p_r \,(p_t)$ are the radial (tangential) in-plane combinations of the $p_x,p_y$-orbitals given by the $\vert p_r (p_t)\rangle=\cos\phi (-\sin\phi)\vert p_x\rangle+\sin\phi (\cos\phi) \vert p_y\rangle$. One may consider that $\vert \vert {\bs k} \lambda \rangle$ should have been used in this work instead of $\vert {\bs k} \lambda \rangle$. Although this is principally correct, is has been studied before that the $\vert \vert {\bs k} \lambda \rangle$ does not change the spin or the spin-momentum orthogonality at the single-particle Hamiltonian level\cite{RevModPhys.82.3045,PhysRevLett.103.266801}. 
 
Determination of the constants $a_0(\bs k,\bs k^\prime)$ and $c_0(\bs k,\bs k^\prime)$ in Eq.(\ref{t_0.1}) with their full momentum dependence is a fundamentally important problem. 
Experimentally, the quasiparticle interference (QPI) with spectroscopic STM can be a promising probe of spin-orbit scattering\cite{PhysRevB.95.115307,PhysRevB.80.245439}. With this technique the authors in Ref.\cite{PhysRevB.95.115307} estimated $\bar c/k_F^2 \simeq 80 \AA^2$ for the polar semiconductor $BiTeI$. Here, the relation between electron-impurity scattering and the spin texture provides an alternative method of extracting $\bar c$ when the warping anisotropy is absent. For a system with inversion symmetry, $\bar c$ can be found once the $\delta_{\lambda k}$ of the spin texture could be measured by using S-ARPES. We know that $\delta_{+ k}\simeq - 20^\circ$ in the case of 
$\alpha$-$Sn$\cite{deviation_angle_alpha_Sn} and the warping  is nearly absent in the surface bands. Using Fig.\ref{isotropic_SE_vs_V0_vs_xnu1} and this $\delta_{+ k}$, we find that $\bar c/k_F^2 \simeq - 40\AA^2$ putting this material as a strong topological spin-orbit impurity scatterer.                                

In summary, we showed that the presence of the spin-orbit scatterings from non-magnetic impurities, an effect which is expected to be finite when the impurities are present in strongly spin-orbit coupled realistic materials, can provide a mechanism for the deviations from the well-established phenomenon of OSML to the one with a non-orthogonal locking. The NOSML angle which can be measured experimentally, is a non-linear function of the impurity concentration and we find that its appearance requires a critical spin-orbit strength. It will be interesting to explore this new state experimentally in more general topological/non-topological systems at various spin-orbit coupling strengths and impurity concentrations. Our theory should pave the road for the full investigation of the effect of the impurities on the spin-momentum locking also including the magnetic impurities. We end with a final remark that, our study highlights additional richnesses of spin textures brought by the impurity effects in strongly spin-orbit coupled materials.   
\section{Acknowledgement}
T.H.’s research is supported by the ITU-BAP project TDK-2018-41181. He dedicates this work to the $250^{th}$ birthday of the Istanbul Technical University (est. $1773$) where a major part of this work was done. T.H. also thanks Northeastern University for support during his visit. The work at Northeastern University was supported by the US Department of Energy (DOE), Office of Science, Basic Energy Sciences Grant No. DE-SC0022216 (accurate modeling of complex magnetic states) and benefited from Northeastern University’s Advanced Scientific Computation Center and the Discovery Cluster and the National Energy Research Scientific Computing Center through DOE Grant No. DE-AC02-05CH11231.  
\appendix
\section{IMPURITY AVERAGING}
\begin{figure}[t]
\includegraphics[scale=0.5,angle=0]{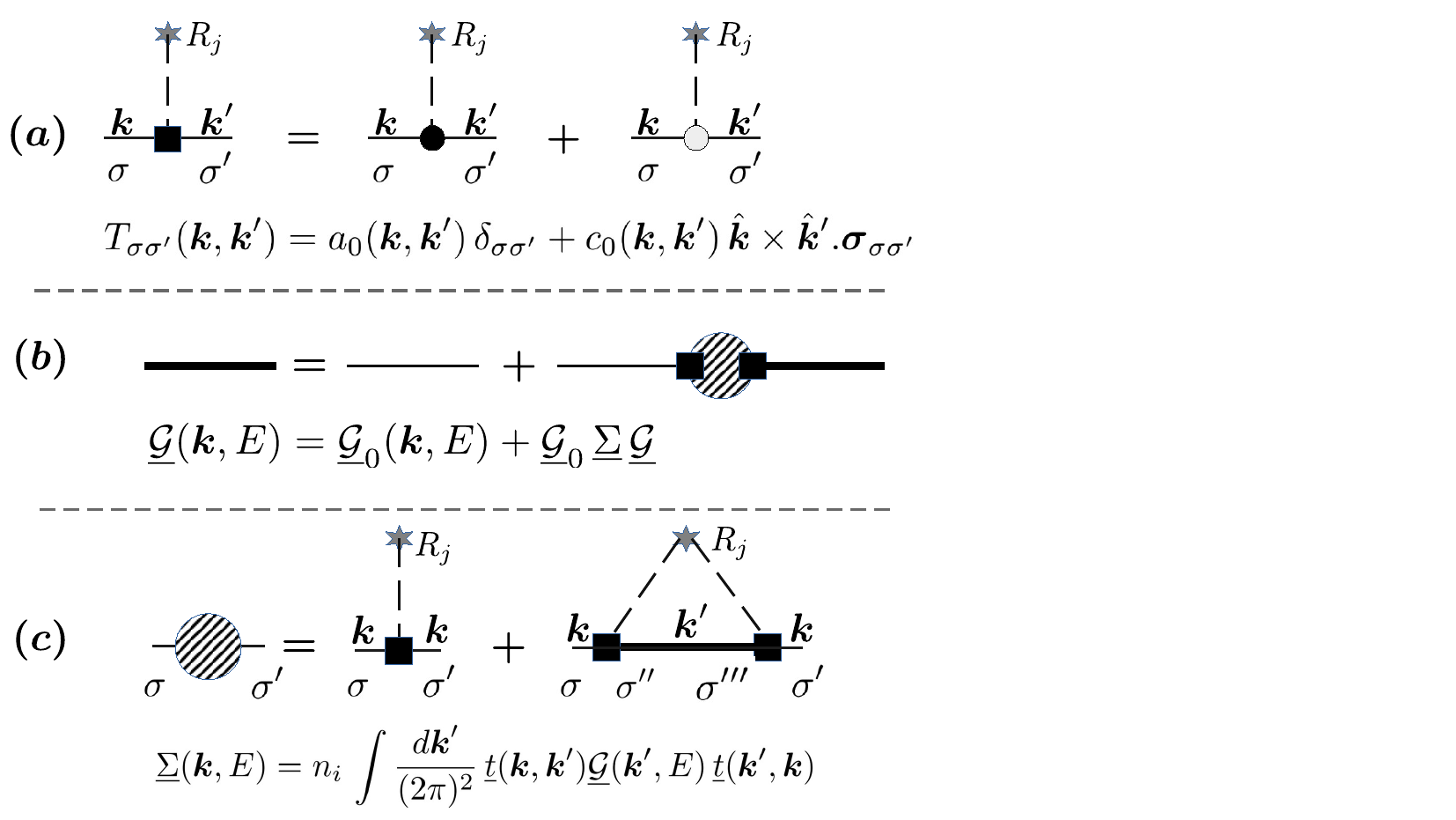} 
\caption{Feynman diagrams corresponding to the first order electron-impurity vertex in Eq.(\ref{t_0.1}) (a), the Green's function in the matrix form (b), the electron self-energy in the matrix form (c).} 
\label{Feynman_1}
\end{figure}

Here we discuss details of the impurity averaging of the electron self-energy and the Green's function. Diagrammatically the electron-impurity interaction is described by the Feynman diagrams as shown in Fig.(\ref{Feynman_1}).
 
We consider that the impurity at the random position $\bs R_j$ is scattered by electrons with initial and final momenta $\bs k, \bs k^\prime$. By the impurity averaging we mean a two-step process. The first is that the kinetic phase $e^{i(\bs k-\bs k^\prime).\bs R_j}$ of the electron wavefunction acquired at the $j$'th scattering is randomized by the random position $\bs R_j$ of the $j$'th impurity. This leads to the average over the impurity positions as described in a separate section below. The second crucial factor is that the random impurity positions also lead to randomized incidence direction of the electron between two scattering events. In order to avoid averaging over the random initial-final momentum orientations at each scattering, we must form scalar quantities of the self-energy vector as $\Sigma_k=\bs \Sigma .\hat{\bs k}$ and $\Sigma_g=\bs \Sigma .\hat{\bs g}_{\bs k}$ as the component of the self-energy along the momentum and along the spin-orbit vector. The $\Sigma_k$ and $\Sigma_g$ are these scalar quantities which are not affected by the random directions of the initial state vector $\bs k$ before each scattering.    

We define the average over the impurity positions by 
\be
\langle O\rangle_{imp}=\int d \bs R \, O(\bs R)\,P(\bs R)
\lb{ia_1}
\ee
Here $P(\bs R)$ is the classical distribution of the impurity positions and $O(\bs R)$ is a generic quantity to be averaged. In our case the impurity positions are completely random with $P(\bs R)=1/\Omega$ with $\Omega$ being the area in which the impurities are randomly scattered. 
 
Considering that the impurity-electron interaction is weak, we use a perturbative expansion of the electron Green's function including the first and second order terms in the impurity-electron scattering matrix elements $T_{\ssp}(\bs k,\bs k^\prime)$. In this section we derive the impurity averaged full self-energy  
given by the Eq.(\ref{Sigma_matrix_1}) of the main text. The latter is given by,
\be 
&&\langle {\underline{\Sigma}(\boldsymbol K)\rangle_{imp}}=\langle {\underline T}(\bs k,\bs k)\rangle_{imp}
\nonumber \\ 
&+&\int \frac{d\boldsymbol k^\prime}{(2\pi)^2}\,  
\langle \, {\underline T}(\bs k,\bs k^\prime) \underline{\cal G}(\boldsymbol k^\prime,E)\,{\underline T}(\bs k^\prime,\bs k)\rangle_{imp} \nonumber \\
\lb{App_Sigma_matrix_1}
\ee
where ${\underline T}(\bs k,\bs k^\prime)$ is the same as the Eq.(\ref{T_kkp}) in the matrix form. 
The Feynman diagrams corresponding to the $T_\ssp(\bs k ,\bs k^\prime)$ are shown in Fig.(\ref{Feynman_1}.a). In Eq.(\ref{App_Sigma_matrix_1}) the $\underline{\cal G}_{\lambda^\prime}(\boldsymbol k^\prime,E)$ is the interacting electron Green's function of the $\lambda^\prime$ band in terms of the $2\times 2$-matrix form in the electron-pseudospin space. In order to find this quantity, we first start with the matrix Dyson equation 
\be 
\frac{1}{\underline{\cal G}({\bs k},E)}=\frac{1}{\underline{\cal G}_0({\bs k},E)}-{\underline \Sigma}({\bs k},E)
\lb{Dyson_eq_1}
\ee
with  
\be 
\underline{\cal G}_0({\bs k},E)=\frac{1}{E-\xi_{\bs k}-\mathfrak{g}_{\bs k}.{\bs \sigma}}
\lb{gee_0}
\ee
representing the non-interacting Green's function 
and the ${\underline \Sigma}({\bs k},E)=\Sigma_0({\bs k},E) \sigma_0+\bs \Sigma({\bs k},E).\bs \sigma$ the full electron self-energy. The $\underline{\cal G}(\bs k,E)$ in the Eq.(\ref{Dyson_eq_1}) can then be compactly written as 
\be 
\underline{\cal G}({\bs k},E)=\sum_\lambda \underline{\cal G}_\lambda({\bs k},E)
\ee
where
\be 
\underline{\cal G}_\lambda({\bs k},E)=\frac{1}{2} [1+\lambda \hat{\bs G}_\lambda(\bs k,E).{\bs \sigma}]\,{\cal G}_\lambda({\bs k},E)
\lb{electron_GF}
\ee 
with  
\be 
{\cal G}_{\lambda}({\bs k},E)=\frac{1}{E-E_{\lambda k}}
\lb{spin-orbit_GF}
\ee
as the exact Green's function of the quasiparticles in the eigenband $\lambda$ of the Hamiltonian in Eq.(\ref{linear_spect}). The ${\bs G}(\bs k,E)=\mathfrak{g}_{\bs k}+\bs \Sigma(\bs k,E)$ is the renormalized spin-orbit vector and $\hat{\bs G}$ is the unit vector of ${\bs G}$. Eq.(\ref{electron_GF}) is the direct sum of the contributions from each spin-orbit band singled out by the physical pole-position of the ${\cal G}_{\lambda}({\bs k},E)$ at $E=E_{\lambda k}$.

We now leave the Green's functions aside and examine the full self-energy in Eq.(\ref{App_Sigma_matrix_1}) diagrammatically in order to derive the dependence of Eq.(\ref{Sigma_matrix_1}) on the impurity concentration. The first term $\langle T_\ssp(\bs k,\bs k)\rangle_{imp}$ is the impurity average of the Eq.(\ref{T_kkp}) for which we use: 
\be 
\sum_{j=1}^{N_{imp}} \langle e^{i(\bs k-\bs k^\prime).\bs R_j}\rangle_{imp}
&=& \frac{1}{\Omega } \int d^{3}\bs R 
\sum_{j=1}^{N_{imp}} e^{i(\bs k-\bs k^\prime).\bs R} \nonumber \\
&=& n_{i}\, \delta_{\bs k,\bs k^\prime}
\lb{App_Sigma_matrix_30}
\ee 
where the average impurity concentration is given by $n_{i}=N_{imp}/\Omega$. We therefore have that $\langle T_\ssp(\bs k,\bs k^\prime)\rangle_{imp}=n_i \delta_{\bs k,\bs k^\prime} t_\ssp(\boldsymbol k,\boldsymbol k)$ which can be ignored since it implies the absence of scattering on the average. We now shift our attention to the second term in Eq.(\ref{App_Sigma_matrix_1}). This requires the knowledge of the full Green's function. The result is (temporarily omitting some indices for simplicity), 
\be 
\langle T \underline{\cal G}(\bs k^\prime,E)\,T^* \rangle_{imp} & \simeq &\frac{1}{\Omega}\sum_{i,j} \langle e^{i(\bs k-\bs k^\prime).(\bs R_i-\bs R_j)} \rangle_{imp} \nonumber \\
& \times & {\underline t}(\bs k,\bs k^\prime)\,
\underline{\cal G}(\bs k^\prime,E) {\underline t}(\bs k^\prime,\bs k) \nonumber \\
\lb{App_Sigma_matrix_2}
\ee 
where we used ${\underline t}(\bs k,\bs k^\prime)={\underline t}^\dagger(\bs k^\prime,\bs k)$ for the unitarity of the scattering matrix. We now work on the relevant part in Eq.(\ref{App_Sigma_matrix_2}) which depends on the impurity average. By definition  
\be 
\sum_{i,j} \langle e^{i(\bs k-\bs k^\prime).(\bs R_i-\bs R_j)} \rangle_{imp}&=& \frac{1}{\Omega}\sum_{i=j} 1 \nonumber \\ 
&+&\sum_{i\ne j} \langle e^{i(\bs k-\bs k^\prime).(\bs R_i-\bs R_j)} \rangle_{imp} \nonumber \\
\lb{App_Sigma_matrix_3}
\ee 
The impurity averaging over a totally random impurity distribution yields random interference between different impurities when $\bs R_i \ne \bs R_j$ yielding a vanishing contribution for $\bs k^\prime \ne \bs k$. This term is therefore (with $1 \ll N_{imp}$)
\be 
\sum_{i\ne j} \langle e^{i(\bs k-\bs k^\prime).(\bs R_i-\bs R_j)} \rangle_{imp}=n^2_{imp} \, \delta_{\bs k,\bs k^\prime}
\lb{App_Sigma_matrix_4}
\ee
Hence it averages out to zero when $\bs k \ne \bs k^\prime$ like the first order impurity average in Eq.(\ref{App_Sigma_matrix_30}). The net effect of this term is therefore essentially the same as the first order impurity-vertex. The net effect of the impurity averaging in Eq.(\ref{App_Sigma_matrix_3}) is therefore provided by the first term on the right hand side as $(1/\Omega)\sum_i 1=N_i/\Omega=n_{i}$. Eq.(\ref{App_Sigma_matrix_2}) is therefore given by 
\be 
\langle T \underline{\cal G}\,T \rangle_{imp} =n_{i} \, {\underline t}(\bs k,\bs k^\prime)\,\underline{\cal G}(\bs k^\prime,E)\,{\underline t}(\bs k^\prime,\bs k)
\nonumber \\
\lb{App_Sigma_matrix_5}
\ee 
Using this result in Eq.(\ref{App_Sigma_matrix_1}) we find 
\be 
\langle {\underline{\Sigma}(\bs K)\rangle_{imp}}= n_i\,\int \frac{d\boldsymbol k^\prime}{(2\pi)^2}\, 
{\underline t}(\bs k,\bs k^\prime) \,\underline{\cal G}(\boldsymbol k^\prime,E) {\underline t}^\dagger(\bs k,\bs k^\prime) \nonumber \\
\lb{App_Sigma_matrix_1_1}
\ee
The Eq.(\ref{App_Sigma_matrix_1_1}) is the full electron self energy corresponding to the Eq.(\ref{SE_1}) in the manuscript. Using Eq.(\ref{trace_ids}), the Eq.(\ref{App_Sigma_matrix_1_1}) yields the Eq.(\ref{Sigma_matrix_1}) in the manuscript where we dropped the explicit impurity averaging symbol $\langle  ... \rangle_{imp}$. 
\begin{figure}[t]
\includegraphics[width=0.5\textwidth]{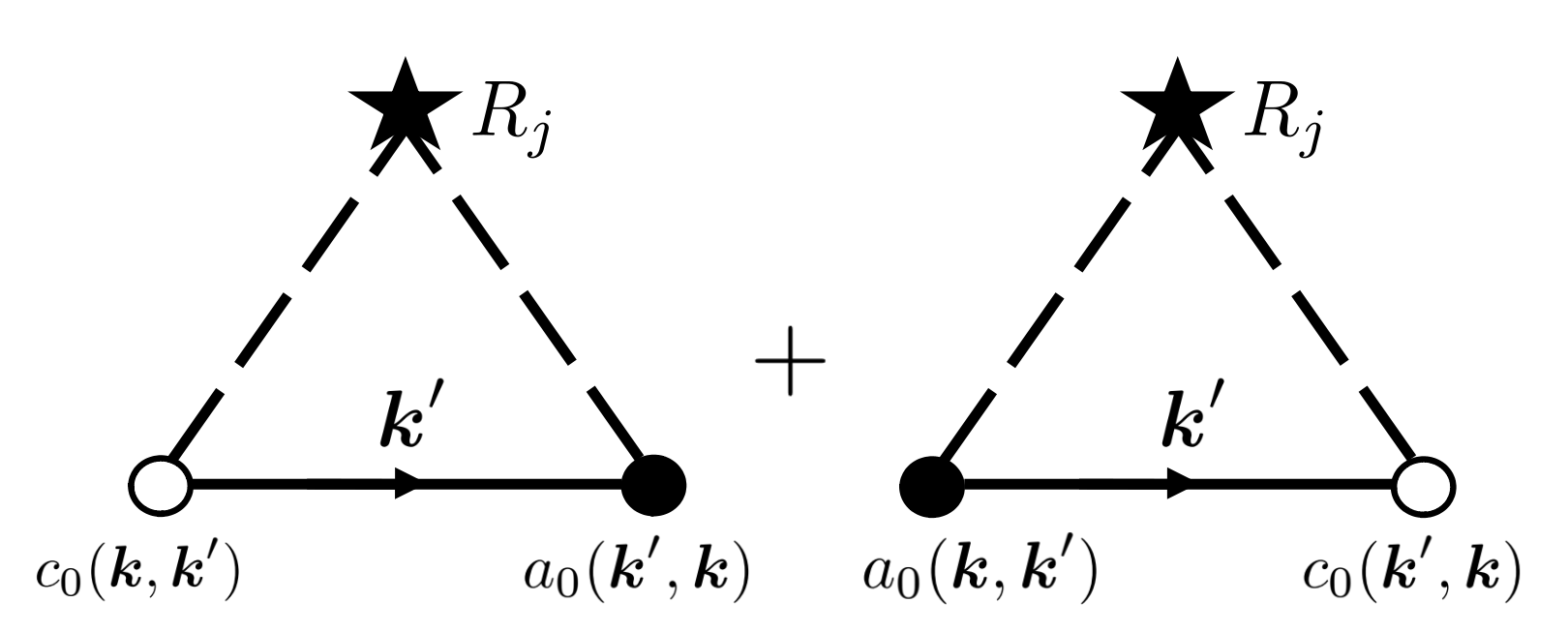}
\caption{Second interference diagrams for the electron self-energy contributing to the NOSML which have linear dependence on the spin-orbit scattering strength $c_0$. The solid line represents the bare electron propagator, and the dashed line represents the two scattering events with the impurity $R_j$.} 
\label{fig:feynman}
\end{figure} 

Next we consider the second type of average which is due to the random orientations of the initial/final momenta. The $\Sigma_k=\langle {\bs \Sigma}.\hat{\bs k}\rangle_{imp}$ and $\Sigma_g=\langle {\bs \Sigma}.\hat{\bs g}_{\bs k}\rangle_{imp}$ are meaningful quantities for impurity averaging since both are scalars and unaffected by the random directions of the scattered electron momenta. It can be explicity seen that, the transformation in Eq.(\ref{complex_SE}) separates the random orientation of the $\hat{\bs k}$ and $\hat{\bs k}^\prime$ by separating out $\phi-\phi^\prime$ in the angular average, and indeed, what remains is the $C_k=\Sigma_g+i\,\Sigma_k$ which is perfectly a scalar complex function of $k$.  

In order to obtain a self consistent expression for $C_k=\Sigma_g-i\,\Sigma_k$, we apply the Eq.'s(\ref{trace_ids}), (\ref{polar_SE}) and (\ref{complex_SE}) in Eq.(\ref{App_Sigma_matrix_1_1}). The real and imaginary parts of $C_k$ define a coupled set of equations given by ($\Sigma_-=\Sigma_x-i\Sigma_y, \, \sigma_-=\sigma_x-i \sigma_y$)  
\be 
\Sigma_-(\bs k,E)=&&\frac{n_i}{2} \int \frac{d\boldsymbol k^\prime}{(2\pi)^2}\, \sum_\lambda\, {\cal G}_\lambda(\bs k^\prime,E) \, t_\mu(\bs k,\bs k^\prime) t_\nu(\bs k^\prime,\bs k) \, \nonumber \\
&&\times \, \frac{1}{2}\,Tr\{\sigma_- \,\sigma_\mu (1+\lambda \bs G(\bs k^\prime,E).\bs \sigma) \sigma_\nu\} ~~~~~~~~ 
\lb{sigma_minus1}
\ee
where $\mu,\nu=0,x,y,z$ and $t_\mu(\bs k,\bs k^\prime)$ refers to the scattering matrix ${\underline t}(\bs k,\bs k^\prime)=t_\mu(\bs k,\bs k^\prime) \, \sigma_\mu$. We further assume that the coefficients $a_0$ and $c_0$ in $t_\mu(\bs k,\bs k^\prime)$ explicitly depend on the scattering angle $\Lambda=\phi-\phi^\prime$\cite{landau2013quantum}. We then use the Eq.(\ref{complex_SE}) on both sides of this expresion and carry out the angular integrations for $\Lambda=\phi-\phi^\prime$  to obtain an expression for $C_k$. Since $\hat{\bs k} \times \hat{\bs k}^\prime=\sin\Lambda \, \hat{\bs z}$, the scattering matrix is confined to those terms with $\mu, \nu=0,z$ in Eq.(\ref{t_0.1}). Applying this in Eq.(\ref{sigma_minus1}), with $G_-=G_x-i G_y=e^{-i(\phi+\pi/2)}\,g_0 k+\Sigma_-$ and the Eq.(\ref{complex_SE}) for $\Sigma_-$,
\be
C_k=\int\frac{k^\prime dk^\prime}{2\pi}\,z_0(k,k^\prime)\,F(k^\prime)~.
\lb{sigma_minus4}
\ee
Here,
\be 
z_0&=&\frac{\operatorname{Im}{\Sigma_0}}{4m}\,\int\frac{d\Lambda}{2\pi}e^{i\Lambda}\,\Bigl [ (t_0 t_0^\prime -t_z t_z^\prime) + (t_0^\prime t_z-t_z^\prime t_0) \Bigr ] \nonumber \\
\lb{z0_1}
\ee
and $F(k)$ is given by the Eq.(\ref{g_av}) in the manuscript. In Eq.(\ref{z0_1}) the short notation $t_0^\prime$ and $t_z^\prime$ imply that $t_0^\prime(\bs k,\bs k^\prime)=t_0(\bs k^\prime,\bs k)$ and $t_z^\prime(\bs k,\bs k^\prime)=t_z(\bs k^\prime,\bs k)$. We now use the fact that $t_0(\bs k,\bs k^\prime)$ and $t_z(\bs k,\bs k^\prime)$ in the notation of Eq.(\ref{sigma_minus1}) are respectively given by $a_0(\bs k,\bs k^\prime)$ and $c_0(\bs k,\bs k^\prime) \, \sin\Lambda$ in the Eq.(\ref{t_0.1}). These coefficients have been calculated for the problem at hand as $a_0(\bs k,\bs k^\prime)=A_0+B_0 \cos\Lambda$ and $c_0(\bs k,\bs k^\prime)=iC_0+4\,D_0 \sin\Lambda$ where $A_0,B_0,C_0,D_0$ are real constants. It can be shown that the first paranthesis on the right hand side in Eq.(\ref{z0_1}) contributes to the real part of the $C_k$, whereas the second one is imaginary and contributes to its  imaginary part $\Sigma_k$ rendering the NOSML as an interference effect between the scalar and the spin-orbit impurity scattering. The angular integration in Eq.(\ref{z0_1}) can be done immediately yielding the complex number,  
\be 
z_0=\frac{\operatorname{Im}{\Sigma_0}}{4m}\,(A_0 + i\, D_0)\, B_0
\lb{z0_2}
\ee
which can then be used in the Eq.(\ref{Ti_2_R}).  

\section{THE SCATTERING MATRIX ${\underline t}(\bs k,\bs k^\prime)$}
In order to construct the ${\underline t}(\bs k,\bs k^\prime)$ we start from a general electron-spinless impurity scattering potential   
$V_{ei}(\bs r)$ as the sum of individually localized electron-impurity potentials at each impurity position $\bs R_j$ as 
\be 
V_{ei}(\bs r)=\sum_{j=1}^{N_{imp}}\,v_{ei}^{(j)}(\bs r-\bs R_j)~.
\lb{imp_pot_1}
\ee  
The $v_{ei}^{(j)}(\bs r)$ is a sum of the spin independent and the  spin-orbit scattering potentials $v_0^{(j)}$ and  $v_{so}^{(j)}$, respectively as 
\be 
v_{ei}^{(j)}(\bs r)&=&v_{0}^{(j)}(\bs r)+v_{so}^{(j)}(\bs r) \, ,\qquad {\rm where}\lb{imp_pot_2} \\
v_{so}^{(j)}(\bs r)&=&\bs \sigma.\Bigl[\bs \nabla v_{0}^{(j)}(\bs r)\times \bs p\Bigr] \nonumber 
\ee
We consider only one type of impurity and assume that $v_{ei}^{(j)}$ is the same for all impurities. The general quantum state of the Bloch electrons is given by 
\be 
\Psi_{\bs k \sigma}(\bs r)=e^{i\bs k.\bs r}\,u_{\bs k \sigma}(\bs r)
\lb{quantum_state}
\ee
with $u_{\bs k \sigma}(\bs r)$ carrying information about the orbital symmetries. The scattering amplitude is given by 
\be 
T_\ssp(\bs k,\bs k^\prime)&=& \langle \bs k^\prime \sigma^\prime \vert V_{ei} \vert \bs k \sigma \rangle \nonumber \\
&=& \int\,d{\bs r} \, \Psi_{\bs k^\prime \sigma^\prime}^*(\bs r) V_{ei}(\bs r) \Psi_{\bs k \sigma}(\bs r) \,  ~~~~~~~
\lb{ei_scattering_general_1}
\ee 
Inserting Eq.(\ref{imp_pot_1}) into Eq.(\ref{ei_scattering_general_1}), we find, 
\be 
T_\ssp(\bs k,\bs k^\prime)=\sum_j^{N_{imp}} \, e^{i(\bs k-\bs k^\prime).\bs R_j} \, t_\ssp(\bs k,\bs k^\prime)
\lb{ei_scattering_general_2}
\ee
where 
\be 
t_\ssp(\bs k, \bs k^\prime) &=& \, [\tilde{v}_{0}(\bs k,\bs k^\prime)]_\ssp+[\tilde{v}_{so}(\bs k,\bs k^\prime)]_\ssp 
\lb{ei_scattering_general_3}
\ee
with
\be 
[\tilde{v}_{X}(\bs k,\bs k^\prime)]_\ssp &=&\int d{\bs r} \Psi_{\bs k^\prime \sigma^\prime}^*(\bs r) (\bs r) v_{X}(\bs r) \Psi_{\bs k \sigma}(\bs r) ~~~~~~~~ \lb{ei_scattering_general_4}
\ee
where $X=0$ for the spinless and $X=so$ for the spin-orbit scatterings as in Eq.(\ref{imp_pot_2}). Due to the spinless character of the $v_0(\bs r)$ in Eq.(\ref{ei_scattering_general_4}), $[\tilde{v}_{0}(\bs k,\bs k^\prime)]_\ssp=a_0(\bs k,\bs k^\prime)\,\delta_{\sigma \sigma^\prime}$. From the Eq.(\ref{ei_scattering_general_4}) it is easily observed that, for $\bs k, \bs k^\prime \simeq 0$ near the center of the linear dispersion, $a_0(\bs k,\bs k^\prime)$ and $c_0(\bs k,\bs k^\prime)$ are proportional to the Fourier transform of the spinless and the spin-orbit potentials in Eq.(\ref{imp_pot_2}). The Eqs.(\ref{ei_scattering_general_1})-(\ref{ei_scattering_general_4}) comprise a generic derivation of Eq.(\ref{t_0.1}) in the main text. This general formulation can be applied to more specific cases only when the Bloch state in Eq.(\ref{quantum_state}) and the electron-impurity potentials in the Eq.(\ref{imp_pot_2}) are given. 
\bibliography{nosml_biblio2}  
\end{document}